\documentclass[fleqn,usenatbib]{mnras}

\usepackage{xspace} 
\usepackage{tabu}
\usepackage[dvipsnames,table]{xcolor}
\usepackage{hyperref}
\usepackage{graphicx,color}
\usepackage{lscape}
\usepackage{longtable,booktabs}
\usepackage{array}

\usepackage[T1]{fontenc}

\DeclareRobustCommand{\VAN}[3]{#2}
\let\VANthebibliography\thebibliography
\def\thebibliography{\DeclareRobustCommand{\VAN}[3]{##3}\VANthebibliography}

\usepackage{graphicx}	
\usepackage{amsmath}	
\usepackage{amssymb}	

\newcommand{\hi}{\textsc{H$\,$i}\xspace}
\newcommand{\ha}{\text{H$\alpha$}\xspace}
\newcommand{\hh}{\text{H$_{2}$}\xspace}
\newcommand{\co}{\text{CO(2-1)}\xspace}

\newcommand{\mstar}{\ensuremath{\text{M}_\star}\xspace}
\newcommand{\mhi}{\ensuremath{\text{M}_{\rm HI}}\xspace}
\newcommand{\kms}{\ensuremath{\text{km}\,\text{s}^{-1}}\xspace}
\newcommand{\msun}{\ensuremath{\text{M}_\odot}\xspace}
\newcommand{\mk}{MeerKAT\xspace}

\newcommand{\caracal}{\texttt{CARACal}\xspace}

\newcommand{\sofia}{\texttt{SoFiA}\xspace}

\graphicspath{{./}{figures/}}

\title[\mk hunts Jellyfish in A2626]{GASP XXXIX: \mk hunts Jellyfish in A2626}

\author[T. Deb et al.]{
Tirna Deb,$^{1,2}$\thanks{E-mail: tirna1106@gmail.com}
Marc A.W. Verheijen$^{1}$,
Bianca M. Poggianti$^{3}$,
Alessia Moretti$^{3}$,
J.M. van der Hulst$^{1}$,
\newauthor
Benedetta Vulcani$^{3}$,
Mpati Ramatsoku$^{4,5}$,
Paolo Serra$^{4}$,
Julia Healy$^{6}$,
Marco Gullieuszik$^{3}$, Cecilia 
\newauthor
Bacchini$^{3}$,
Alessandro Ignesti$^{3}$,
Ancla Müller$^{7}$,
Nikki Zabel$^{8}$,
Nicholas Luber$^{9,10}$,
Yara L. Jaffé$^{11}$, 
\newauthor
Myriam Gitti$^{12,13}$
\\
$^{1}$Kapteyn Astronomical Institute, University of Groningen, Landleven 12, 9747 AV Groningen, The Netherlands\\
$^{2}$Department of Physics and Astronomy, University of the Western Cape, Robert Sobukwe Road, Bellville 7535, South Africa\\
$^{3}$INAF- Osservatorio Astronomico di Padova, Vicolo dell'Osservatorio 5, I-35122 Padova, Italy\\
$^{4}$INAF- Osservatorio Astronomico di Cagliari, Via della Scienza 5, I-09047 Selargius (CA), Italy\\
$^{5}$Department of Physics and Electronics, Rhodes University, PO Box 94, Makhanda, 6140, South Africa\\
$^{6}$ASTRON, the Netherlands Institute for Radio Astronomy, Postbus 2, 7990 AA Dwingeloo, The Netherlands\\
$^{7}$Ruhr University Bochum, Faculty of Physics and Astronomy, Astronomical Institute, Universitätsstr. 50, 44801 Bochum, Germany\\
$^{8}$Department of Astronomy, University of Cape Town, Private Bag X3, Rondebosch 7701, South Africa\\
$^{9}$Department of Physics and Astronomy, West Virginia University, P.O. Box 6315, Morgantown, WV 26506, USA\\
$^{10}$Center for Gravitational Waves and Cosmology, West Virginia University, Chestnut Ridge Research Building, Morgantown, WV 26505\\
$^{11}$Instituto de Física y Astronomía, Universidad de Valparaíso, Avda. Gran Bretãna 1111 Valparaíso, Chile\\
$^{12}$Dipartimento di Fisica e Astronomia, Università di Bologna, via Gobetti 93/2, 40129 Bologna, Italy\\
$^{13}$INAF, Istituto di Radioastronomia di Bologna, via Gobetti 101, 40129 Bologna, Italy\\
}

\date{Accepted 2022 August 24. Received 2022 August 24; in original form 2021 September 27}

\pubyear{2021}

\begin{document}
\label{firstpage}
\pagerange{\pageref{firstpage}--\pageref{lastpage}}
\maketitle

\begin{abstract}
We present \mk \hi observations of six jellyfish candidate galaxies (JFCGs) in the galaxy cluster, A2626. Two of the six galaxies JW100 and JW103, that were identified as JFCGs from B-band images, are confirmed as jellyfish galaxies (JFGs). Both of the JFGs have low \hi content, reside in the cluster core, and move at very high velocities ($\sim$ 3$\sigma_{cl}$). The other JFCGs, identified as non-jellyfish galaxies, are \hi rich, with \hi morphologies revealing warps, asymmetries, and possible tidal interactions. Both the A2626 JFGs and three other confirmed JFGs from the GASP sample show that these galaxies are \hi stripped but not yet quenched. We detect \hi, \ha, and \co tails of similar extent ($\sim$ 50 kpc) in JW100. Comparing the multi-phase velocity channels, we do not detect any \hi or \co emission in the northern section of the tail where \ha emission is present, possibly due to prolonged interaction between the stripped gas and the ICM. We also observe an anti-correlation between \hi and \co, which hints at an efficient conversion of \hi to \hh in the southern part of the tail. We find that both RPS and \hi-to-\hh conversion are significant depletion channels for atomic gas. \hi-to-\hh conversion is more efficient in the disc than in the tail.


\end{abstract}




\section{Introduction} \label{sec:intro}
In the dense cluster environment, hydrodynamical interactions such as ram-pressure stripping (RPS) between the inter-stellar medium (ISM) of galaxies and intra-cluster medium (ICM) play an important role in transforming galaxies \citep{Gunn1972} from star forming (SF) and gas rich, to quiescent and gas poor. `Jellyfish' galaxies (JFGs) are extreme examples of RPS with `tentacles' of material that stretch tens of kpcs beyond their stellar discs \citep{Smith2010, Fumagalli2014, Ebeling2014}.For example, one of the most studied jellyfish galaxy is ESO137-001 in the nearby Norma cluster, with its long, extended tail observed in multi-wavelength \citep{sun2007, jachym2014, fossati2016, jachym2019}.

The extended, fragile atomic hydrogen (\hi)  gas discs are not only the reservoirs for star formation, but also serve as diagnostic tracers of gravitational and hydrodynamic environmental processes \citep{Bravo2001,chung2009,Jaffe2015}. Images of the \hi\ gas in galaxies show that the extended \hi\ discs are readily affected by the cluster environment, resulting in \hi\ stripped and \hi\ deficient galaxies compared to field galaxies \citep{Verheijen2001, Oosterloo2005,chung2009,Serra2013,Gogate2020,Loni2021}. This deficiency and stripping is less conspicuous for the centrally concentrated and more tightly bound molecular hydrogen gas (\hh). Studies of cluster galaxies often show conflicting results that differ between clusters: low SF in \hi\ stripped but \hh\ normal galaxies (Virgo cluster, \citealt{Kenney1989,Bumhyun2017, brown2021}), and normal or slightly enhanced SF in \hi\ deficient but \hh\ rich galaxies (JFGs, \citealt{Moretti2020b,Moretti2020a}). In the Virgo cluster, some galaxies are \hi\ deficient and show signatures of RPS, but are only moderately deficient in \hh (\citealt{chung2009,Kenney1989, brown2021}). In Fornax, galaxies can be \hi or \hh deficient, or deficient in both gas phases  \citep{Zabel2019, Loni2021}. 

Until recently, observations and studies of the environmental impact on the multi-phase ISM have been limited to nearby clusters such as Virgo (\citealt{chung2009,Kenney1989, brown2021}), Fornax \citep{Zabel2019, Loni2021} or Coma \citep{Boselli1997, Casoli1996, Healy2021a}. It is still not clear what the relative influence of the ICM pressure is on the \hi\ and \hh\ gas. It also remains a puzzle why \hi\ stripped galaxies appear to be quiescent while the \hh\ gas, which is the more direct constituent for SF, is not deficient. There are still many other questions without clear answers: How do environmental processes impact the baryon cycle in galaxies? How does SF quench in galaxies: by gas removal or consumption? Therefore, it is very timely to broaden our understanding of environmental processes acting on the multi-phase ISM of galaxies in clusters. JFGs, extreme examples of RPS and often located in the dense cluster core, are excellent objects to investigate environmental effects on the multi-phase ISM.

The GAs Stripping Phenomena in galaxies survey (GASP; \citealt{Poggianti2017}) was carried out with the Multi Unit Spectroscopic Explorer (MUSE), observing a statistically significant sample of JFGs in nearby clusters (z=0.04-0.07) over a wide range of stellar masses, morphological asymmetries and environments. The key scientific motivation is to investigate how, where and why gas removal occurs, and to quantify the amount of star formation involved in these processes \citep{Poggianti2017}. Within GASP, JFGs are defined as the galaxies with a \ha\ tail at least as long as the diameter of the stellar disc. In the RPS tails of these JFGs, often \textit{in situ} SF occurs by efficient conversion of \hi\ to \hh\ followed by the collapse of molecular clouds due to thermal instabilities, turbulent motion etc \citep{Poggianti2019, Mueller2021}. \hi\ gas is also ionised by the young stars that are formed \textit{in situ} in the tail. \cite{PoggiantiNature2017} observed a strong correlation between RPS and the presence of an Active Galactic Nucleus (AGNs), hinting at the possibility that ram pressure is causing gas to flow towards the centre thus triggering the AGN activity. \cite{poggianti2016}
identified possible stripped galaxies as jellyfish candidates galaxies (JFCGs) purely based on B-band imaging. JO201, JO204, JO206, and JW100 were confirmed stripped galaxies i.e. JFGs with long tails by MUSE observations \citep{Poggianti2017, gullieuszik2017, bellhouse2017, bellhouse2019, Poggianti2019}.

Five of the GASP JFGs (JO201, JO204, JO206, JO194, JO135) were also observed with the Jansky Very Large Array (JVLA; \citealt{Perley2009}) to investigate their \hi\ gas content. \cite{ramatsoku2019, ramatsoku2020} observed that both JO201 and JO206 show a strong stripping of \hi\ gas  relative to the stellar disc (on the sky and/or in velocity). Both the \hi\ and \ha\ tails are of similar extent, and these galaxies have comparable \hi\ deficiencies, with enhanced SF compared to that of galaxies of similar stellar and \hi\ mass. However, it is worth noting that JO201 resides in a more massive and dense cluster than JO206 \citep{bellhouse2017, bellhouse2019, Poggianti2017, jaffe2018}. \cite{Deb2020} observed a much longer (90 kpc) \hi\ tail compared to the RPS \ha\ tail (30 kpc, \citealt{gullieuszik2017}) in JO204.  \cite{Deb2020} also investigated whether RPS is inducing the AGN in JO204 by funnelling \hi\ gas towards the centre of the galaxy, resulting in a redshifted \hi\ absorption profile, but the study was inconclusive. \cite{luber2022} investigated the \hi properties of non-jellyfish galaxies in clusters with JFGs from GASP sample. Our study will add the \hi picture of the of the JFCGs in A2626, and will complement the detailed studies of JW100 in other wavelengths \citep{Poggianti2019}.

\cite{HD2021} have conducted a blind \mk\ \hi\ survey of the galaxy cluster A2626, which hosts six JFCGs including JW100, signifying ongoing environment driven evolution of the galaxies. In this paper, we compare the \hi\ properties and SF of the A2626 JFCGs with other JFGs from the GASP sample residing in different clusters. In the next sections we will first address the \hi\ in all six JF(C)Gs and then concentrate on understanding the details of the RPS process in JW100 for which all ISM phases are now available from \cite{Poggianti2019}, \cite{Moretti2020b} and our MeerKAT observations. \hi\ observations of JW100 are the last piece of the puzzle to complete the picture of the interplay of the multi-phase (atomic, molecular, ionised) ISM with itself and with the ICM.

The paper is organised as follows: Sec. \ref{sec:overview} gives an overview of A2626 and the JFCGs in it. \hi observations and data processing are presented in Sec. \ref{sec:obs}. In Sec. \ref{sec:atlas} we explore the \hi properties of the JFCGs. In Sec. \ref{sec:hidef}, we investigate the relation between \hi deficiency and star-formation rate (SFR) in the JFGs. In sec. \ref{sec:JW100_chnl_maps}, we then focus on JFG JW100 and study the atomic, molecular, and ionised gas phases in different velocity channels. In sec. \ref{sec:hi_dep_chnls}, we calculate the contribution of different depletion mechanisms to the stripping of the multi-phase gas in JW100. Finally in sec. \ref{sec:summary}, we summarise the findings and conclusions.

Throughout this paper, we adopt a Chabrier initial mass function (IMF; \citealt{Chabrier2003}), and assume a cold dark matter cosmology with $\Omega_{M}$ = 0.3, $\Omega_{\Lambda}$ = 0.7 and H$_{0}$ = 70 km s$^{-1}$ Mpc$^{-1}$. At the cluster redshift (z=0.055292, \citealt{HealySS2021}), this yields 1$\arcsec$=1.074 kpc.


\renewcommand{\arraystretch}{1}
\begin{table}
    \centering
    \caption{A2626 cluster properties}
    \begin{tabular}{ll}
    \hline
     \noalign{\vspace{0.5mm}}
     \multicolumn{2}{l}{Cluster properties}                   \\
     \hline
     \noalign{\vspace{0.5mm}}
     RA          & 23:36:31$^{a}$         \\
     Dec   & +21:09:36.3$^{a}$   \\
     z      & 0.055292$^{b}$ \\ 
     $\sigma_{cl}$            & 660 $\pm$ 26 km/s$^{b}$   \\
     M$_{200}$    & 3.9 $\times$ 10$^{14}$ \msun$^{c}$ \\
     R$_{200}$   & 1.59 Mpc$^{b}$                    \\
     L$_{\rm X}$ & 1.9 $\times$ 10$^{44}$ erg/s $^{d}$        \\
     \hline\hline
    \end{tabular}
        \begin{flushleft}
    \footnotesize  $^{a}$ \cite{cava2009}, $^{b}$ \cite{HealySS2021}, $^{c}$ \cite{biviano2017}, $^{d}$ \cite{wong2008}
    \end{flushleft}
    \label{tab:obsparam}
\end{table}

\section{A2626 and its jellyfish candidate galaxies}
\label{sec:overview}

\begin{figure*}
     \includegraphics[width=\textwidth]{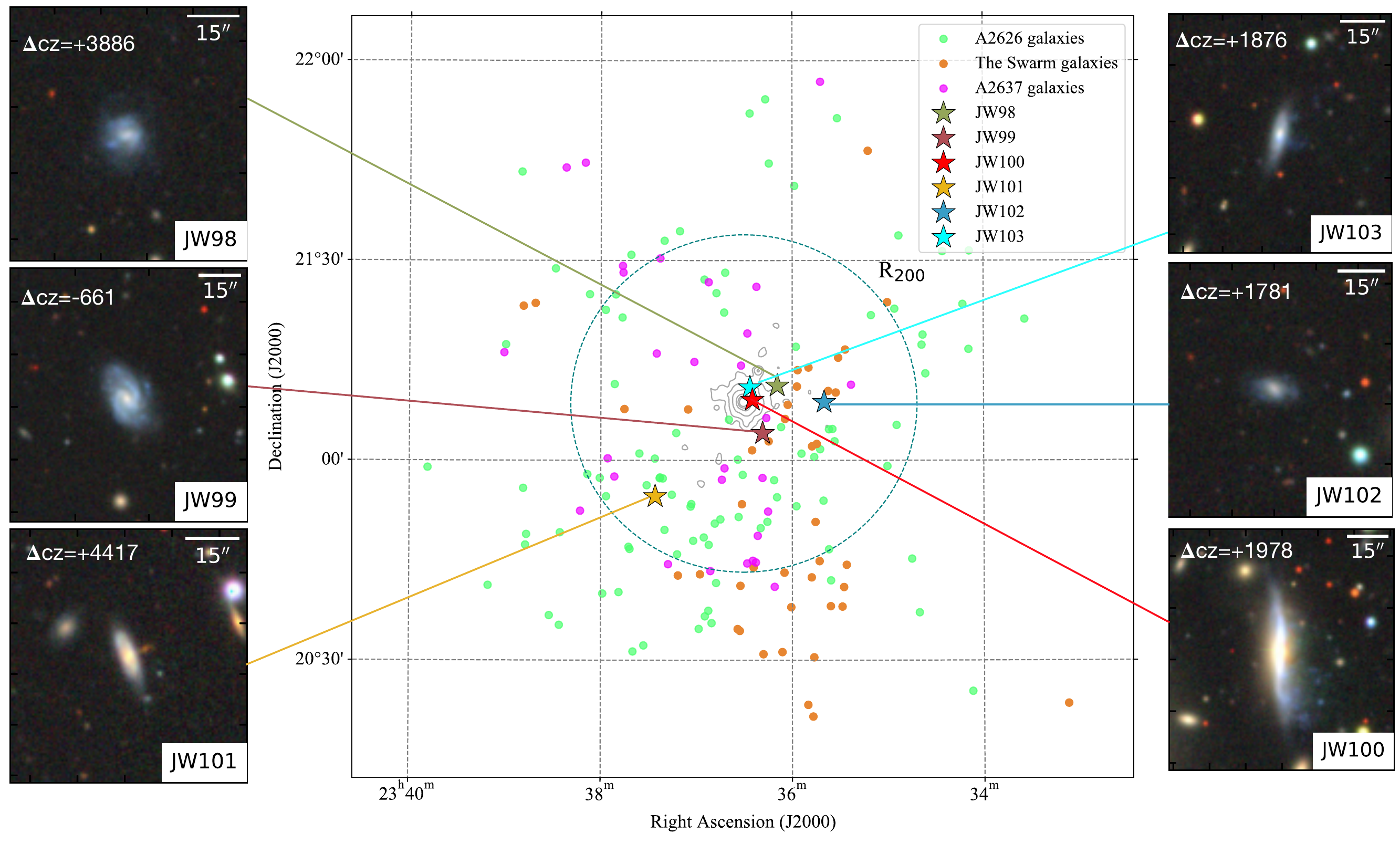} 
    \caption{The location of jellyfish candidate galaxies (JFCGs) JW98, JW99, JW100, JW101, JW102, and JW103 in the A2626 volume \citep{poggianti2016}. The dashed circle represents the R$_{200}$ of A2626. The black contours represent the X-ray emission in A2626 from ROSAT images. The coloured dots represent the \hi detections in different redshift over-densities in the A2626 volume \citep{HealySS2021}. JClass=1 JFCGs (i.e. least probable JFGs from their optical morphologies, JW98, JW99, JW101, JW102) are outside the cluster core, higher JClass (JClass $\geq$ 2) galaxies JW100 and JW103 are residing close to the cluster core in projection. On the side left and right panels we show the optical Dark Energy Camera Legacy Survey (DECaLS, \citealt{dey2019}) images of JFCGs in A2626, including the $\Delta$cz velocity difference with respect to the cluster core.}%
    \label{fig:JFCG_sky_dist}%
\end{figure*}

\subsection{The galaxy cluster A2626}

The focus of our study are the JFCGs identified in A2626, one of 77 clusters imaged as part of the WIde-field Nearby Galaxy-cluster Survey (WINGS: \citealt{Fasano2006}). A2626 is a moderately massive, X-ray emitting cluster at a redshift of z=0.055292 \citep{HealySS2021}. The global properties of the cluster are listed in Table 1. It is a cooling-core cluster \citep{wong2008, McDonald2018, Ignesti2018, Kadam2019}. The X-ray surface brightness and thermodynamic maps show cold fronts and spiral wakes, indicate sloshing and the disturbed state of the ICM of A2626 \citep{Rizza_2000, wong2008, McDonald2018, Kadam2019}. The cluster hosts a peculiar central Brightest Cluster Galaxy (BCG, IC 5338) that displays an offset nucleus in an asymmetric stellar envelope. The BCG is a radio-loud galaxy and the source of a peculiar and well-studied radio source known as `The Kite' \citep{Rizza_2000, Gitti_2004, Gitti_2013b, Kale_2017, Ignesti_2017, Ignesti2018, Ignesti_2020b}. Using a new spectroscopic survey, \cite{HealySS2021} identified six galaxy groups/substructure within and beyond the projected R$_{200}$ radius. These substructures indicate that A2626 is actively accreting galaxies and galaxy groups from its surroundings, and confirm that A2626 is a dynamically active cosmic environment.

\begin{table*}
    \renewcommand{\arraystretch}{1.1}
    \centering
    \caption{Properties of the jellyfish candidate galaxies}
    \begin{tabular}{lcccccccccc}\hline
        \multicolumn{1}{c}{Name}    & RA      & DEC   & Optical & $\Delta$cz    & \mstar$^{d, e}$   & SFR$^{c, e}$ & $\pm^{c, e}$ & \mhi\  &$\pm$   & JClass$^{a}$ \\
        & (J2000)$^{a}$ & (J2000)$^{a}$  &  redshifts$^{b,c}$  & (\kms) & $\times$ 10$^{9}$ & \msun\/yr$^{-1}$ & \msun\/yr$^{-1}$ & $\times$ 10$^{9}$ & $\times$ 10$^{9}$ &   \\ 
          & deg & deg  &         &   & \msun    &      &   &  \msun          &     \msun                \\ \hline
        JW98  &   23:36:09.41  &  +21:11:11.8  & 0.0681  & +3886     &   5.2$^{+2.6}_{-1.7}$ &      0.31       &   0.17               & 11   & 0.4  &   1       \\
       JW99          &    23:36:18.55  &  +21:04:04.4    &       0.05327    &   -661      &   8.8$^{+3.4}_{-2.5}$      &    2.7     &  0.9      & 24    & 0.5  &        1          \\
       JW100    &    23:36:25.01 &   +21:09:03.6  & 0.06189 &  +1978   &    320$^{+310}_{-120}$     &          4           &       0.8       & 2.8   & 1.2   &  5        \\
        JW101     &   23:37:25.92  &  +20:54:35.3 & 0.06983 &  +4417  &  16$^{+7}_{-5}$       &   4.2              &    1.7             &  8.9 &  0.7     &    1      \\
        JW102    &    23:35:40.01  &  +21:08:44.5   &  0.0613 &  +1781    &  2.7$^{+48}_{-2.6}$       &  0.3   & 0.18   &  6.5  & 0.3   &   1      \\
        JW103    &   23:36:26.54 &   +21:10:55.6 & 0.06189 &  +1876     &    5.2$^{+4.4}_{-2.8}$    &   2.2  & 0.9  &  1  & 0.3   &    2      \\
         \hline
    \end{tabular}
    \begin{flushleft}
    \footnotesize  $^{a}$\cite{poggianti2016}, $^{b}$\cite{HealySS2021} , $^{c}$\cite{Poggianti2019},  $^{d}$ \cite{vulcani2018},  $^{e}$ based on WISE, provided via private communication
    \end{flushleft}
    \label{tab:JFCG_prop}
\end{table*}

\subsection{Jellyfish candidate galaxies in A2626}
\label{sec: JFCGs}

 \begin{figure*}
   \centering
    \includegraphics[width=1.0\textwidth]{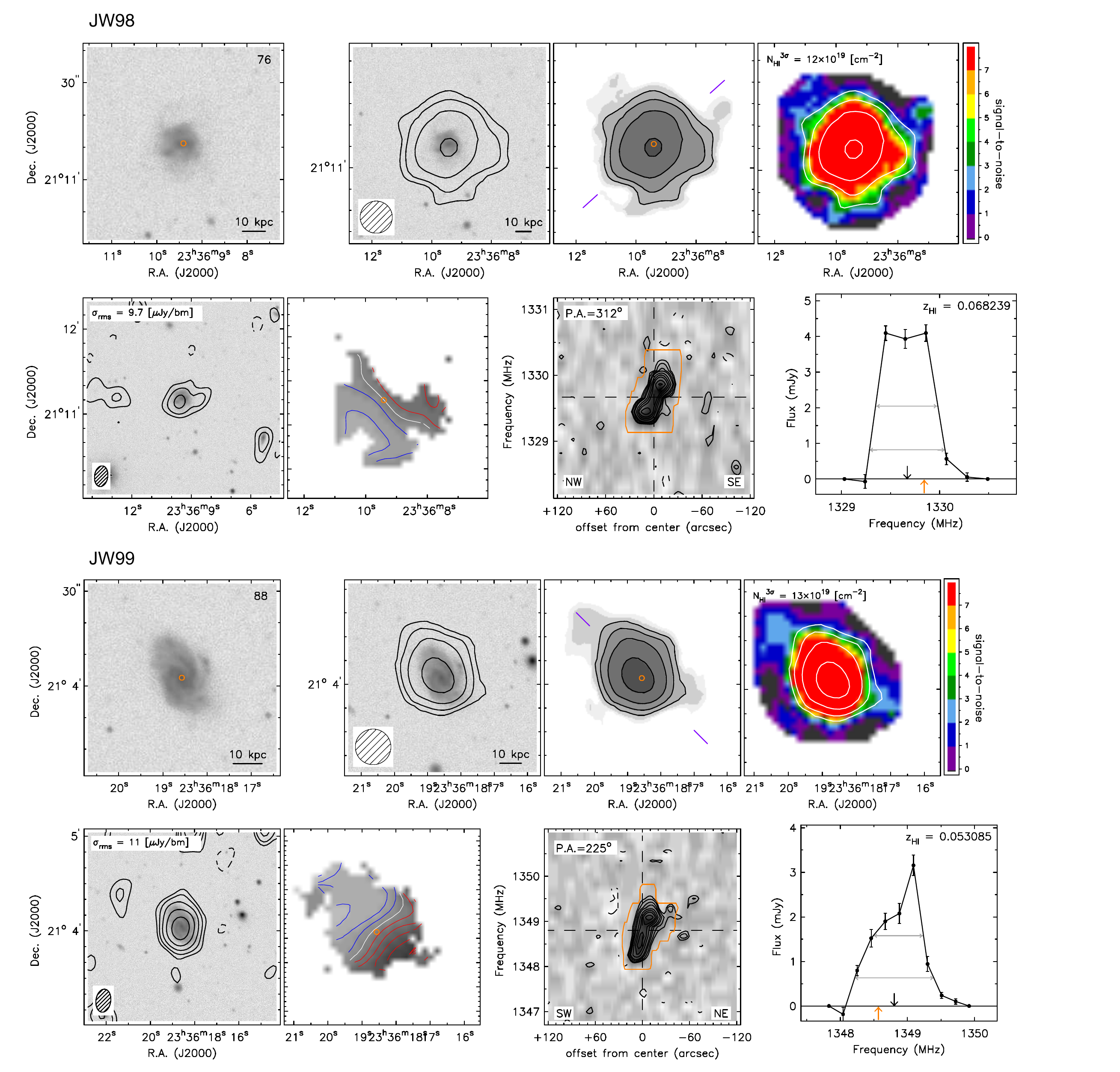}
    \vspace{-12pt}
    \caption{Atlas pages for JW98 and JW99. Clockwise from the top left panel: DECaLS r-band image, the \hi\ column density contours on DECaLS g-band image in greyscale, total \hi\ map in greyscale, signal-to-noise map as a colour map with the different colours representing different levels of signal-to-noise, \hi\ global profile, \hi\ position-velocity (PV) diagram, \hi\ velocity field, and radio continuum map on  DECaLS g-band image. The \hi\ column density contours are plotted at the levels N$^{3\sigma}_{\hi} \times 2^{n}$ with n=0,1,2,3,...  with N$^{3\sigma}_{\hi}$ - the average ${3\sigma}$ column density for each galaxy. The \hi\ beam size (15") is shown in the bottom left. In the velocity field, the lighter greyscales and blue contours show the approaching side with contour levels 25, 50, 75, 100,..200 \kms. The systemic velocity v$_{sys}$ is shown by the white contour. The darker greyscales and red contours represent the receding side of the \hi\ rotating disc with the contour levels -25, -50, -75, ..., -200 \kms. The radio continuum contours are plotted at the levels 2$^{n}$ of the RMS noise, where n=0,1,2,3,... Dashed contours are drawn at -2$\sigma$ the RMS noise. The hatched elliptical beam in the radio continuum map is the beam size of continuum map (14.6" $\times$ 9.30", 171.6$^{\circ}$). The position-velocity (PV) diagrams are made along the major axis of the optical image, starting at the receding side of the galaxy. The position angles are mentioned on the top of each panel which are determined by visual inspection of the optical image while the receding side with respect to the systemic recession velocity of the galaxy is determined from the three-dimensional \hi\ datacube. The directions of the start and end points of the PV slices are indicated in the bottom left and right corners of each panel respectively. The orange contour outlines the manually made optimum mask. The vertical dashed line indicates the optical centre of the galaxy from the SDSS. The horizontal dashed line corresponds to the central frequency as derived from the global profile. The \hi\ redshift is calculated from this central frequency and reported in the top right of the bottom right panel. The \hi\ spectrum or global profile of each galaxy is constructed by applying the dilated, optimum \hi\ mask for each galaxy to the 20" resolution \hi\ datacube. The vertical black downward arrow in the middle represents the central frequency, corresponding to the systemic velocity, which is the midpoint of the 20\% velocity width. The measured optical redshifts are shown with an orange upward arrow. The method to derive the errorbars as shown on the global profile is described in Section 5.1 in \protect \cite{HD2021}.}
\label{fig:JW98_JW99_atlas}
\end{figure*}

\begin{figure*}
   \centering
    \includegraphics[width=1.0\textwidth]{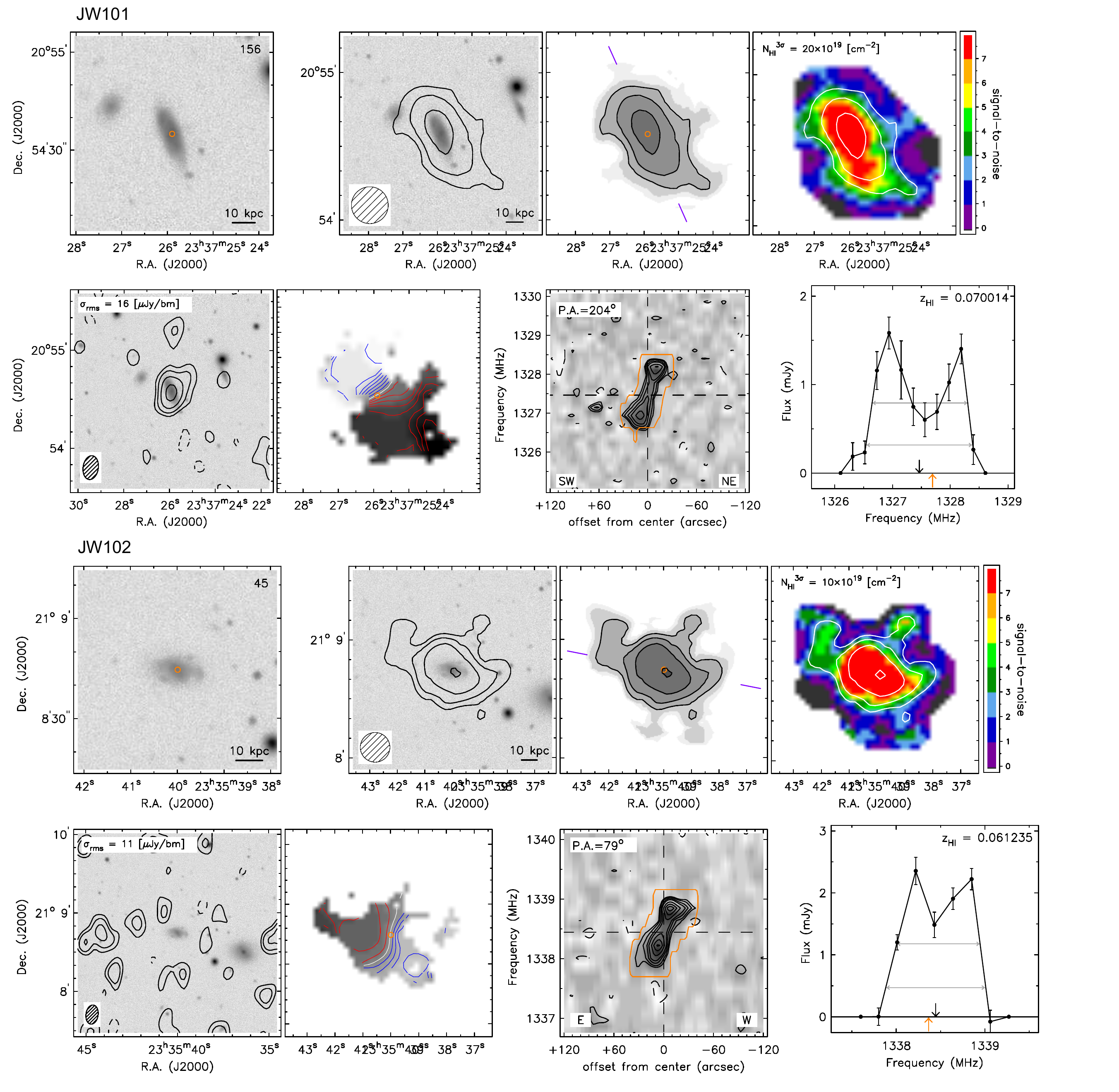}
    \vspace{-12pt}
    \caption{Atlas pages for JW101 and JW102. The description of atlas pages is the same as for Fig. \ref{fig:JW98_JW99_atlas}. } 
\label{fig:JW101_JW102_atlas}
\end{figure*}

A2626 hosts six JFCGs as identified by \cite{poggianti2016}. Using WINGS B-band images to search for optical evidence of gas stripping, the JFCGs were identified based on their optical morphology. \cite{poggianti2016} ranked the galaxies on a scale of 1 to 5, JClass = 1 being the least plausible JFG and JClass = 5 being the most plausible JFG. The six JFCGs identified in A2626 are JW98, JW99, JW100, JW101, JW102, and JW103. Their global properties are listed in Table \ref{tab:JFCG_prop} and illustrated in Fig. \ref{fig:JFCG_sky_dist} which also indicates their projected locations in A2626. Of these galaxies, JW100 has the highest JClass=5 and is known to be experiencing ram-pressure stripping (RPS) of its ISM. In the top right panel of Fig.  \ref{fig:JFCG_sky_dist}, JW100 clearly displays and extended RPS tail towards the South West that appears as light blue in the optical image that indicating extra-planar star formation activity, which was confirmed by MUSE observations \citep{Poggianti2019}. JW100 has been the subject of many detailed studies (as summarised below) while the other five JFCGs were not part of the GASP MUSE sample. JW103 has a low JClass=2 and it is likely at a later stage of RPS than JW100 due to its lower stellar mass. It is at a same velocity and similar location at the cluster as JW100 with two extensions of blue light towards the West in the optical image (see top right panel of \ref{fig:JFCG_sky_dist}). The other four candidate JFGs have JClass=1 and are least likely to be experiencing RPS (see Fig. \ref{fig:JFCG_sky_dist}).

It should be noted that JW100, JW102 and JW103 have a very similar radial velocity offset with respect to A2626 of $\Delta$cz $\approx$+1800 km/s ($\sim 3\sigma_{\rm cl}$) indicating that these galaxies are entering A2626 as a group. JW100 and JW103 with JClass$>$1 are close to the X-ray emitting cluster core while JW102 is still at a substantial projected distance ($\sim$ 0.6 R$_{200}$) and thus not strongly affected by RPS yet. 

The optical morphologies of the JFCGs show a wide variety of characteristics, ranging from irregular blue systems like JW98, which \cite{poggianti2016} suggest to be the result of harassment, to two-armed grand-design spirals with a hint of lopsidedness like JW99, to bulge-dominated like JW101. Notably, the optical image of JW100 shows some reddening of the stellar light just west of its bulge, suggesting that dust is also being removed from the disk of this galaxy. In the lower JClass galaxies, it is not always obvious which blue features hint as ongoing RPS, but their overall disturbed or lopsided optical morphologies classified these galaxies as JFCGs. The spatially resolved \hi\ observations presented in this paper will clarify whether these JClass=1 galaxies are actually experiencing RPS.

Using stellar masses and SFRs determined using WISE photometry (T. Jarrett, private communication), we can compare the A2626 JFCGs to other galaxies in A2626 and different reference samples (see Sec. \ref{sec:hidef}). The stellar masses are calculated from the 3.4$\rm \mu$m (W1) and 4.6$\rm \mu$m (W2) emission. The SFRs are determined from the PAH emission in the 12$\rm \mu$m (W3) band; see \cite{cluver2014, cluver2017, jarrett2013, jarett2019} for details.

All the above mentioned studies of JFCGs in A2626 lack observational information on an important constituent of the ram-pressure stripped gas, namely the \hi gas, and that is what our current study contributes. \hi\ observations of JFCGs in A2626 will provide a diverse picture of \hi\ morphologies of such galaxies that are at different stages of stripping and can confirm whether candidate JFGs are actually experiencing ongoing RPS or not. Particularly, JW100 is one of the most well studied  JFGs so far and information about the atomic gas will complete the picture of the multi-phase gas of JW100.

\section{\hi observations and data processing}
\label{sec:obs}

A2626 was observed with the \mk\ telescope \citep{Jonas2016} as one of the first \mk-64 open time observations (project code SCI-20190418-JH-01). A2626 was observed during the nights of 15-17 July 2019 for a total of 15 (3 $\times$ 5) hours. The large field-of-view of \mk\ (61$\arcmin$ $\times$ 61$\arcmin$ at a frequency of 1346 MHz where \hi\ emission from z = 0.055 is detected) is adequate to cover the cluster beyond R$_{200}$ and encompasses all the JFCGs. The sensitivity of \mk\ enables a 3$\sigma$ \hi\ mass detection limit of 2 $\times$ 10$^{8}$ \msun\ in the field centre with a linewidth of 300 \kms, and a corresponding 5$\sigma$ column density of  2.5 $\times$ 10$^{19}$ cm$^{-2}$ at an angular resolution of 30$\arcsec$. The data calibration and imaging was carried out using \caracal, (version 1.0, previously known as \texttt{MeerKATHI}, \citealt{Jozsa2020}) to reduce the data and \sofia-2 \citep{Serra2015} to detect the \hi\ sources within the \mk\ cube. For the complete details on the data calibration, imaging, and source finding, see \cite{HD2021}. 

The four low class (JClass = 1) candidate galaxies turned out to be gas rich and were identified by \sofia-2 (\hi\ ID=76 (JW98), 88 (JW99), 156 (JW101), 45 (JW102) in \citealt{HD2021}). The atlas pages for JW98 and JW99, and JW101 and JW102 are shown in Fig. \ref{fig:JW98_JW99_atlas} and \ref{fig:JW101_JW102_atlas} respectively. JW103 (JClass=2) and JW100 (JClass=5) were not identified by \sofia-2 with the general parameter settings used in \cite{HD2021}. Both galaxies have very faint \hi\ emission, and could only be identified visually with the knowledge of their coordinates and redshifts from optical observations. The data were smoothed using Gaussian kernels to angular resolutions of 15\arcsec, 20\arcsec, 25\arcsec, 30\arcsec, and with boxcar kernels to velocity resolutions of $\sim$135, $\sim$225, and $\sim$315 \kms to enhance the sensitivity of the data. An angular resolution of 20\arcsec\ and a velocity resolution of $\sim$135 \kms resulted in the highest signal-to-noise \hi\ emission for both JW100 and JW103. Masks were constructed manually for each velocity channel containing \hi\ emission with an \hi\ flux of at least three times the rms noise in that channel, and visually detecting spatial coherence in the distribution of the \hi\ emission compared to the neighbouring channels. We have then \texttt{CLEAN}-ed \citep{Hogbom1974} the `dirty' channels of the 20\arcsec\ \hi\ datacube using the mask made, down to the 0.3$\sigma$ level. The \texttt{clean} components were \texttt{restore}d with a 20\arcsec circular Gaussian beam. The pixels in the \texttt{clean}ed 20\arcsec\ datacube that are outside the mask were set to zero and pixels inside the mask were summed up along the part of the frequency axis contained within the mask to obtain the total \hi\ maps for JW100 and JW103.


\section{\hi properties of the jellyfish candidate galaxies}
\label{sec:atlas}

\begin{figure*}
   \centering
    \includegraphics[width=1.0\textwidth]{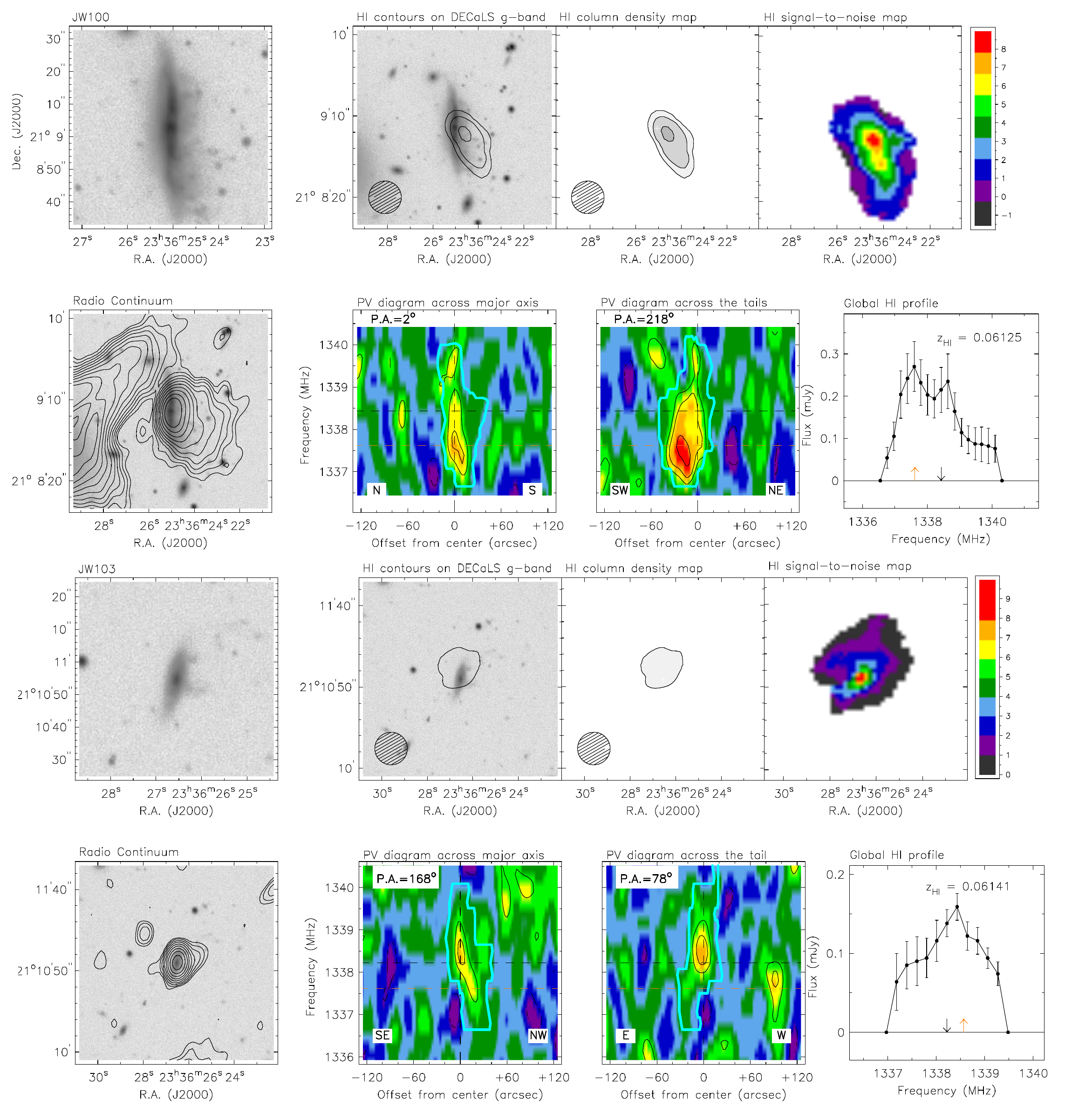}
    \vspace{-12pt}
    \caption{Atlas pages for JW100 and JW103. All the data products except the second and third panel in the bottom row are the same as in Fig. \ref{fig:JW98_JW99_atlas}. The second and third panels of the second row are a \hi Position-Velocity (PV) diagram along the optical major axis and a \hi position-velocity (PV) diagram across the tail respectively. The \hi\ column density contours are plotted at the levels N$^{3\sigma}_{\hi} \times 2^{\rm n}$ with n=0,1,2,3,...  while N$^{3\sigma}_{\hi}$, the average ${3\sigma}$ column density for each galaxy and the \hi\ beam size (20\arcsec) is shown in the bottom left of the \hi maps. The cyan contour in the PV diagrams outlines the manually made optimum mask as described in the section \ref{sec:obs}.} 
\label{fig:JW100_atlas}
\end{figure*}

The \hi\ morphologies of the gas-rich JFCGs JW98, JW99, JW101, and JW102 do not reveal any obvious stripped jellyfish-like gas tails at the angular resolution of our MeerKAT observations  (15$\arcsec$ corresponding to 16 kpc at the redshift of A2626, see Fig \ref{fig:JW98_JW99_atlas} and Fig \ref{fig:JW101_JW102_atlas}). The detailed description of the methods used to derive the \hi\ properties of JW98, JW99, JW101, and JW102 can be found in \cite{HD2021} and the description of atlas pages of these galaxies are in Fig \ref{fig:JW98_JW99_atlas} and \ref{fig:JW101_JW102_atlas}. JW98 and JW101 are at higher redshifts (see Table \ref{tab:JFCG_prop}) than the redshift range of A2626 (0.0475 $<$ z $<$ 0.0615, \citealt{HealySS2021}) and thus are not likely members of the cluster. According to \cite{poggianti2016}, JW98 is possibly a harassed galaxy, the radio continuum shows an extension to the west which may support the harrassment scenario. JW99 is a somewhat lopsided galaxy, both in the optical and in \hi, and displays a mild kinematic asymmetry (see the velocity field in Fig. \ref{fig:JW98_JW99_atlas}). JW101 and JW102 both seem to have warped \hi\ discs, likely due to interactions with their neighbouring galaxies. All of these JFCGs have significant \hi\ masses ($\sim$10$^{10}$ M$_{\odot}$), often exceeding their stellar mass. 

 Fig. \ref{fig:JW100_atlas} shows the \hi\ properties of the JFGs JW100 and JW103 in the form of atlas pages. Due East of JW100, at the centre of A2626, is the peculiar, extended radio continuum source known as the ``kite” source. For both JW100 and JW103, the optical and HI redshifts are slightly offset (\hi\  redshift is within the error of the optical redshift). \mk\ \hi\ observations of JW100 reveal a striking \hi\ tail extending south-westwards up to $\sim$ 20 kpc from the nucleus of the galaxy. Most of the \hi\ emission associated with JW100 is in the tail, which is evident in the PV diagram along the tail (see Fig. \ref{fig:JW100_atlas}). For JW103 the signal-to-noise of the \hi\ emission is very low and we do not observe any \hi\ tail in the direction of the optical tentacles which are visible in the B-band image. However, the \hi emission is mostly on one side of the galaxy (north) and blue-shifted in JW103. The total \hi\ masses of JW100 and JW103 (see equation 5 in \citealt{HD2021}) seen in emission are 2.8$\times$10$^{9}$ \msun\ and 1$\times$10$^{9}$ \msun, respectively. Interestingly, in both JW100 and JW103, we observe an uni-directional radio-continuum tail in the direction of the optical tail. The radio tail is produced by synchrotron emission of cosmic rays accelerated in the disc that have been advected in the tail by the ram pressure wind \citep{ignesti2022}. Evidence of non-thermal radio tails have been frequently reported in jellyfish galaxies (e.g. \citealt{Chen_2020, Roberts_2021a}).
 
 So, with the combination of \hi and optical morphologies of the six the JFCGs in A2626, we infer that only JW100 is an actual JFG while JW103 is a ram-pressure stripped galaxy at a later stage of stripping. Other JFCGs namely JW98, JW99, JW101, and JW102 are identified as non-jellyfish galaxies from their \hi observations. They are \hi rich  (\mhi $\sim$10$^{10}$ M$_{\odot}$) and do not show signs of RPS, but reveal mildly disturbed \hi morphologies consistent with harassment or tidal interactions.


\section{\hi deficiency and SFR enhancement in jellyfish galaxies}
\label{sec:hidef}

The relation between stellar masses of galaxies and star formation rates (the `star formation main sequence'; SFMS) is a well-known scaling relation (e.g. \citealt{brinchmann2004, noeske2007, elbaz2007, speagle2014, tomczak2016}). We investigate the location of the JFGs and JFCGs with respect to the SFMS. For that purpose, we use the scaling relation from \cite{cluver2020}. Both the SFMS and quenching threshold relations in \cite{cluver2020} are calibrated using the the WISE data and the same methodologies that is used for stellar mass and SFR calculations for our sample galaxies.

\begin{figure*}
     \includegraphics[width=1.0\textwidth]{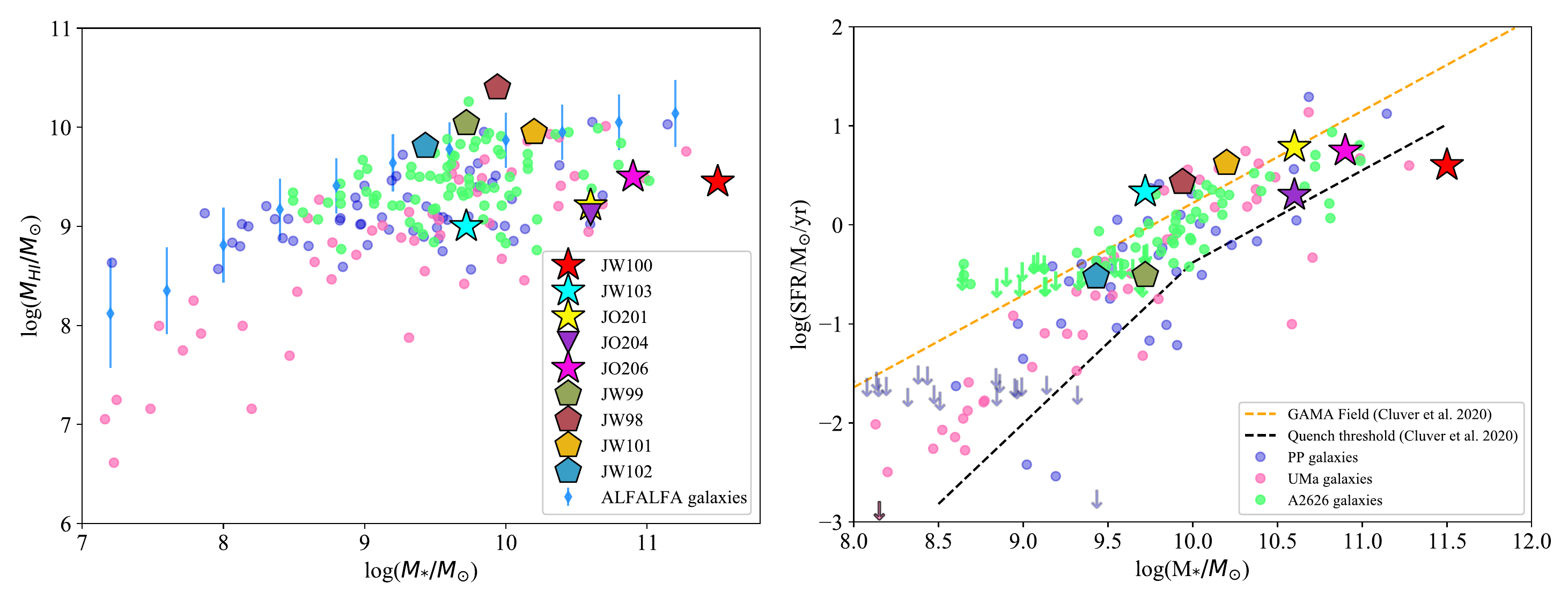} 
    \caption{Left panel: \hi\ mass as a function of stellar mass for jellyfish candidate galaxies (JFGs and JFCGs) compared to reference sample galaxies. The field galaxies are the  ALFALFA sample from \citealt{maddox2015}, the other coloured circles are \hi\ detected galaxies from the PP, UMa and A2626 volumes. The star markers are identified as JFGs in our \mk sample and the previous JVLA studies of GASP galaxies \protect \citep{ramatsoku2019, ramatsoku2020, Deb2020}. The pentagon markers are jellyfish candidate galaxies (JFCGs) selected from optical B-band images which are identified as non-jellyfish galaxies from their \hi\ morphologies. Clearly, the JFGs have lower \hi\ content compared to field galaxies or other reference sample galaxies at fixed stellar mass. Right panel: SFMS for JFGs and JFCGs. The SFMS relation and quenching threshold are taken from \protect \cite{cluver2020} which is calibrated using the WISE data and the same methodologies that is used for stellar mass and SFR calculations for our sample galaxies. JFGs are scattered around the SFMS. The downarrows are 2${\sigma}$ upper limits on the SFR from WISE. }
    \label{fig:Mstar_SFR_HI}
\end{figure*}

    \begin{figure}
     \includegraphics[width=0.5\textwidth]{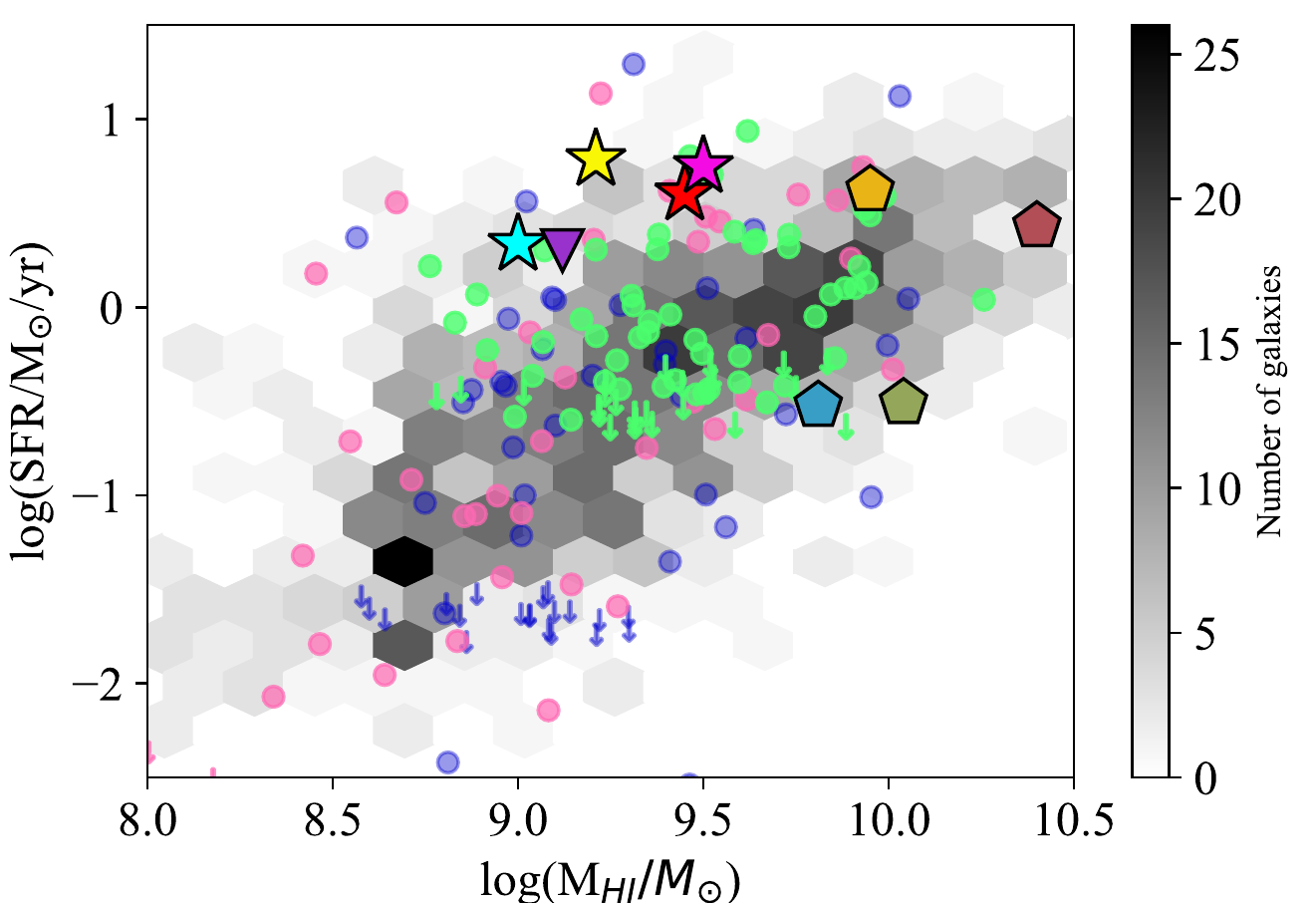} %
    \caption{SFR vs \hi mass for JFGs and JFCGs compared with the reference sample galaxies with the same stellar mass range. Grey hexagons are galaxies from the xGASS sample \citep{catinella2010} and the star, and pentagon markers represent the same JFGs and JFCGs galaxies as in Fig. \ref{fig:Mstar_SFR_HI}. JFGs are \hi deficient but their star formation is not quenched yet.}%
    \label{fig:MHI_SFR}%
\end{figure}

We have also included the galaxies detected in the volume-limited \hi\ surveys of the Ursa Major (UMa, \citealt{Verheijen2001, busekool2021}), Perseus-Pisces (PP, Bilimogga et al. in prep), and the entire A2626 volumes as reference samples. The UMa volume is defined as a `supergroup' that lacks any central concentration while still being gravitationally bound \citep{wolfinger2013}. At a distance of 18.6 Mpc, the volume is characterised by mostly late-type galaxies with a normal gas content \citep{tully1977}. We use the \hi\ detections from a blind \hi\ imaging survey of the volume with the JVLA-D configuration (for more details see \citealt{busekool2021}) supplemented by targeted observations of 50 UMa galaxies (described in \citealt{Verheijen2001}). The PP volume is a section of a ridge embedded in the PP supercluster. A blind \hi\ imaging survey of this volume has been performed with the JVLA-C configuration (Bilimogga et al. in prep). At a distance of 66 Mpc, the volume consists of several loose groups of galaxies that follow the large-scale structure of the PP supercluster (see \citealt{Trasarti1998}). Since the UMa and PP observations are also volume limited \hi\ surveys like the \mk A2626 survey, we used those galaxies to compare with our JFCGs. For A2626, 97 \hi\ sources other than JW100 and JW103 are detected in our \mk\ obervations (see \citealt{HD2021}), and they are also used in this paper as a reference sample. Stellar masses and SFRs for PP and UMa galaxies have also been calculated from WISE photometry using the same prescription as for the A2626 galaxies as mentioned in Sec. \ref{sec: JFCGs}.

The left panel of Fig \ref{fig:Mstar_SFR_HI} shows \hi\ vs stellar masses for the JFCGs and the reference sample galaxies. Three other \hi\ detected JFGs namely JO201, JO204, and JO206 \citep{ramatsoku2019, ramatsoku2020, Deb2020} from the GASP sample \citep{Poggianti2017} are also plotted with JW100 and JW103 in coloured star markers. Four non-stripping JFCGs are plotted with pentagon coloured star markers. Together with PP, UMa, and A2626 galaxies, we have also plotted the results from the Arecibo Legacy Fast ALFA- Sloan Digital Sky Survey (ALFALFA–SDSS) galaxy sample of field galaxies with errorbars from \cite{maddox2015}, which spans a large range in stellar and \hi\ masses. All five galaxies that are under the influence of RPS (JO201, JO204, JO206, JW100, and JW103) have relatively low \hi\ masses for their stellar masses when compared to the reference sample galaxies (also found by \citealt{luber2022}). Especially, when compared with the ALFALFA field galaxies, all five JFGs are almost 1 dex below the typical \hi\ masses of the galaxies of similar stellar masses. 

The right panel of Fig \ref{fig:Mstar_SFR_HI} shows SFR vs stellar masses for the JFGs, the reference sample galaxies and the SFMS relation. For all but JW103, the SFRs are measured with MUSE \citep{vulcani2018}. The SFR calculations from MUSE and WISE observations are verified to be consistent (within $\sim$ 10\% scatter). For JW103, we use the SFR calculated from the 12$\mu$m (W3) emission. Interestingly, both JFGs and JFCGs which are later identified as non-jellyfish galaxies do not stand out with respect to the SFMS of \cite{cluver2020}, i.e. they do not necessarily have enhanced SFR compared to their stellar masses, this is contrary to what is found by \cite{vulcani2018}. However, we note that JW100 lies below the quenching threshold as defined by \cite{cluver2020}, signifying a strongly reduced star formation activity for its stellar mass.

Finally, we inspect how the atomic gas content is linked to the global star formation activity of the JFGs as compared to usual field or cluster galaxies following Fig 6 in \cite{ramatsoku2020}. We have prepared a reference sample of field galaxies with the same stellar mass range as the five JFGs (9.7 $\leq$ log$M_{*}$ $\leq$11.5) from the Extended GALEX Arecibo SDSS Survey (xGASS; \citealt{catinella2010}). Fig. \ref{fig:MHI_SFR} displays SFR vs  M$_{\rm HI}$ for the xGASS sample \citep{doyle2006, huang2012, saintonge2016} in addition to the PP, UMa and A2626 cluster galaxies. The coloured star and pentagon markers represent the JFGs and other JFCGs respectively. Intriguingly, all the five JFGs lie above the distribution of the reference samples. In other words, the JFGs have higher star formation rates compared to their \hi\ masses for galaxies with similar stellar masses \citep{ramatsoku2019, ramatsoku2020}. This suggests that these galaxies are being \hi stripped but are not yet quenched. Furthermore, this also implies that the little \hi\ gas these galaxies retain, is more efficiently transformed into \hh gas, as concluded by \cite{Moretti2020a}, resulting in a high SFR for their \hi masses, compared to normal galaxies of similar \hi\ masses. Given that these JFGs are \hi\ deficient and often do not have enhanced SFR for their \mstar, they must have evolved from right to left in Fig. \ref{fig:MHI_SFR}, and not so much from bottom to top. We note in the right panel of Fig. 5, that the JFGs do not lie significantly above the SFMS and therefore, do not currently have an enhanced SFR for their stellar mass. Therefore, we conclude that the JFGs have not moved upwards significantly in Fig. 6. However, we cannot entirely rule out that a mild enhancement of the SFR may have occurred during a short period of time since these galaxies entered the clusters.


\section{The Multi-phase ISM of JW100}
\label{sec:JW100_chnl_maps}

\subsection{Previous studies on JW100}

JW100 is the proto-typical JFG. It is a nearly edge-on spiral galaxy with an AGN in its centre \citep{Poggianti2019, radovich2019}. JW100 resides in A2626 under optimal conditions for ram-pressure stripping \citep{jaffe2018, gullieuszik2020}, with a particularly high line-of-sight velocity of 2000 \kms with respect to the cluster rest frame and a projected distance of 83 kpc from the cluster centre \citep{Poggianti2019}. It is the most massive JFG in the entire GASP sample with a stellar mass of 3.2 $\times$ 10$^{11}$ \msun \citep{Poggianti2019}. JW100 is one of the most spectacular JFGs with extra-planar RPS tails of ionised and molecular gas, UV stellar light, X-ray and radio continuum emission \citep{Poggianti2019, Moretti2020b, ignesti2022}.

 In the northern part of the tail, there is a lack of extra-planar clumpy CO gas and star formation while only diffuse H$\alpha$ is observed, with a high [OI]/H$\alpha$ ratio and X-ray emission. This suggests that the excess \ha\ emission in the northern part of the tail is possibly a result of a complex ICM-ISM interplay triggered by RPS (e. g. \citealt{ferland2009, fossati2016, cramer2019, campitiello2020}). We note here that the northern, diffuse \ha emission has a recession velocity that is inconsistent with the rotational velocity of JW100, suggesting that the ionised gas is decelerated along the orbit and longer exposed to interaction with ICM (see Sec. \ref{sec: multiphase_interplay} for more details). The radio continuum emission from the tail of JW100 is mainly synchrotron emission caused by relativistic electrons ejected by supernovae explosions, indicating the presence of magnetic fields pervading the stripped gas (as in e.g. \citealt{Chen_2020, Mueller2021, Roberts_2021a}). 

Based on ALMA observations, \cite{Moretti2020b, Moretti2020a} found a large amount of molecular gas ($\sim$ 2.5 $\times$ 10$^{10}$ \msun), i.e. 8\% of the stellar mass, a value more than an order of magnitude higher than that found in typical Virgo cluster galaxies \citep{corbelli2012, brown2021} and in the local xCOLD GASS sample \citep{saintonge2017} for galaxies of similar stellar mass. The ALMA data reveal a 35 kpc long RPS tail of clumpy and diffuse molecular gas from the centre of JW100 \citep{Moretti2020b}. The high values of the CO line ratio (r$_{21}$ $\geq$ 1) in the southern part of the tail, the narrow, single-component CO emission lines, and low velocity dispersion hint that the tail in that region is composed of dense star forming molecular gas. This signifies that these clouds are formed in situ, either from the stripped \hi or the diffuse molecular gas  \citep{Moretti2020b, Moretti2020a}.

\subsection{The distribution of \hi in JW100 in comparison with H${\alpha}$ and CO}
\label{sec: multiphase_interplay}

\begin{figure}
 {%
    \includegraphics[width=.45\textwidth]{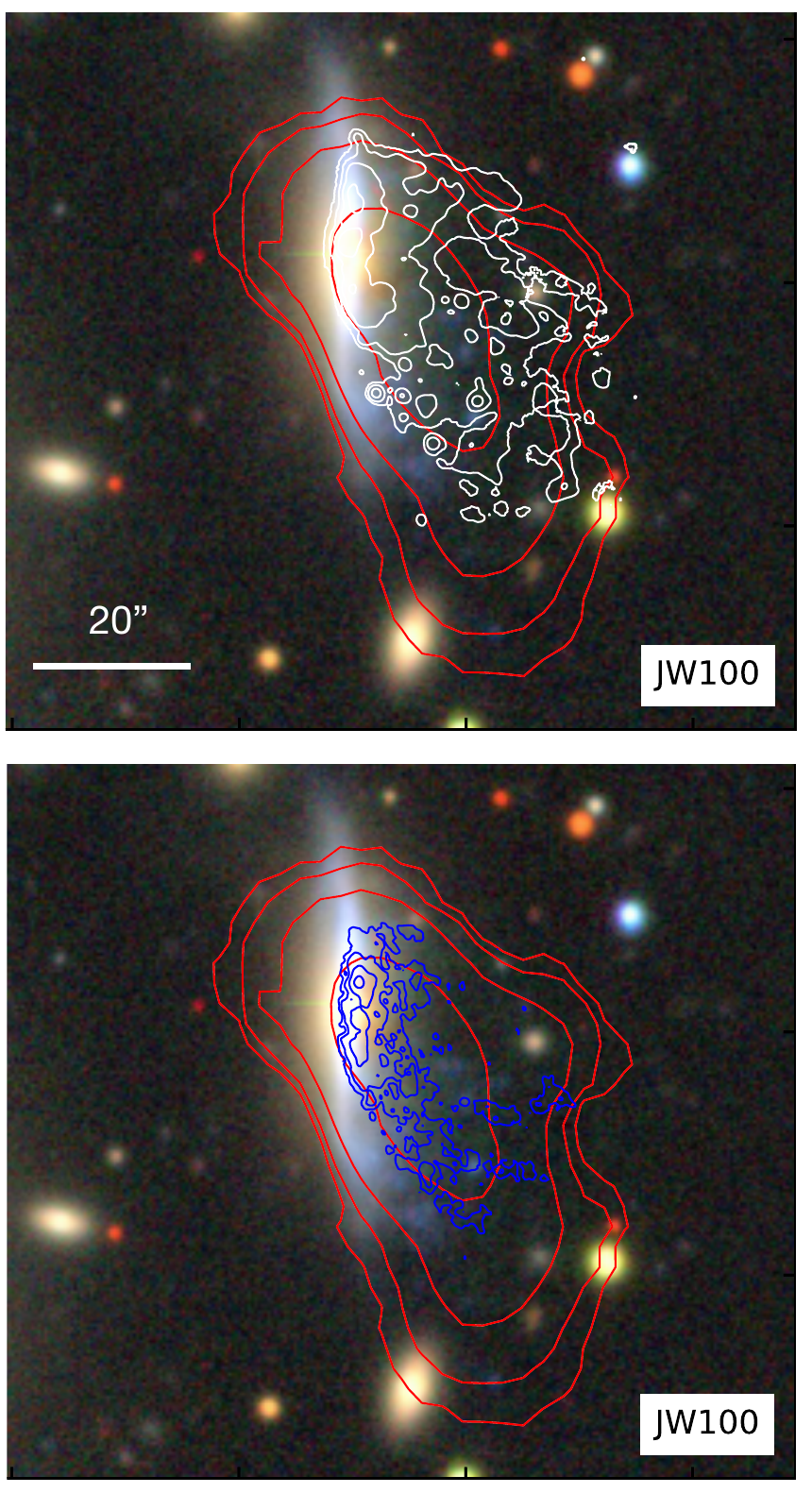} 
  } 
  \caption{Multi-phase gas in the RPS tail in JW100. Top panel: \ha (white, 1$\arcsec$ resolution) and \hi (red,  20$\arcsec$ resolution) contours overlayed on the DECaLS color image. Bottom panel: \co (blue,  1$\arcsec$ resolution) and \hi (red,  20$\arcsec$ resolution) contours overlayed on DECaLS color image. 1$\arcsec$ corresponds to 1.074 kpc at a distance of JW100. } 
  \label{fig:HI_HA} 
 \end{figure}

\begin{figure*}
    \centering
    {\includegraphics[width=\textwidth]{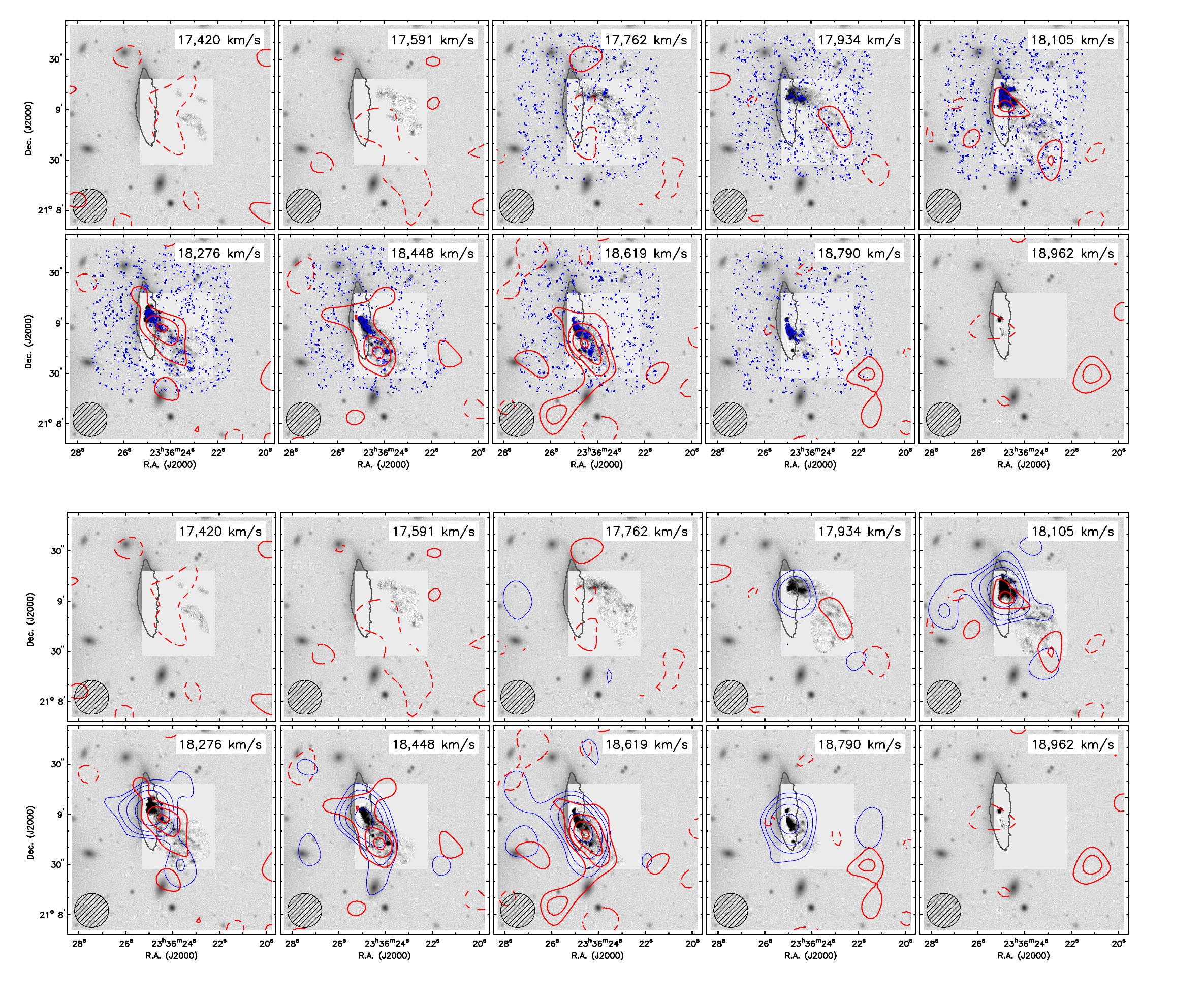}}
   \caption{\ha, \co, and \hi\ channel map overlays for JW100. In the bottom two rows, the \co\ emission is smoothed to the same angular resolution as the \hi\ emission (20$\arcsec$) to compare the atomic and molecular gas phases on the same spatial scales.  The red solid contours represent the \hi\ emission with contour levels of (2, 3, 4, 5,..) $\times \sigma$ or RMS and the dotted red contour represents the $-2\sigma$ level. The blue contours indicate \co\ emission with contour levels (2, 4, 8, 16,..) $\times \sigma$ where $\sigma$= 0.28 mJy/beam for the high resolution ALMA cube and 0.67 mJy/beam for the 20$\arcsec$ ALMA cube. The central light-grey rectangle represents the area in which \ha\ emission is observed with MUSE, after substraction of the [NII] emission lines and the stellar continuum emission. The image is larger than the field of view of MUSE, so the DECaLS g-band image is shown in the background. The black contour represents the stellar disc and the hatched circle on the bottom left of each panel shows the size of the 20$\arcsec$ \hi\ beam. \ha\ emission is seen in the northern part of the RPS tail but no \co or \hi emission while both \ha\ and CO are anti-correlated with \hi on small scales.}
    \label{fig:chnl_maps}
\end{figure*}

Fig. \ref{fig:HI_HA} shows the multi-phase gas in the RPS tail of JW100. The top and bottom panel of Fig. \ref{fig:HI_HA} show the \hi emission (red,  20$\arcsec$ resolution) on \ha (white,  1$\arcsec$ resolution) and \hi emission (red,  20$\arcsec$ resolution) on \co (blue,  1$\arcsec$ resolution) respectively with the DECaLS \citep{dey2019} color image in the background. The atomic, molecular, and ionised gas tails are almost of similar extent ($\sim$ 50 kpc), this small difference of the tail lengths could be also due to the different sensitivities and resolutions of the \hi, \ha, and \co data. 

We then inspect the distribution of gas at different recession velocities through channel maps in Fig. \ref{fig:chnl_maps}.  We compare the \hi\ contours (red), \co\ contours (blue), and the \ha\ emission (greyscale) at different velocities in JW100 in ten panels. The top two rows of Fig. \ref{fig:chnl_maps} show \ha\ emission at the original MUSE angular and velocity resolution (1$\arcsec$ $\times$ 1$\arcsec$, 57.1 \kms, \citealt{Poggianti2019}), \co\ emission at the original ALMA resolution (1.4$\arcsec$ $\times$ 1.1$\arcsec$, 44.1 \kms, \citealt{Moretti2020b}) and \hi\ emission smoothed to 20$\arcsec$ and $\sim$ 135 \kms. The high resolution ALMA channel maps indicate the location of small \co\ clumps. In the bottom two rows of Fig. \ref{fig:chnl_maps}, the \co\ emission is also smoothed to an angular resolution of 20$\arcsec$ to compare the atomic and molecular gas phases on the same spatial scales. 

\ha\ emission is visible already in the top-left panels (17,420-17,762 \kms) where \hi\ and \co\ are not yet visible. \co\ and \hi\ emission are detectable from the 4th and 5th panel respectively. In the 6th, 7th, and 8th panel of Fig. \ref{fig:chnl_maps}, \hi\ emission is found further down the tail (away from the centre of JW100) than the \co\ clumps. In the bottom two rows of Fig. \ref{fig:chnl_maps}, though the \co\ clumps are smoothed in those panels, the centres of the \co\ emission are still offset from the \hi\ emission. The 3$\sigma$ \hi\ `appendix' at the bottom left in the 8th panel, the 3$\sigma$ \hi\ peaks in the 9th and 10th panels, and the -2$\sigma$ \hi\ peaks in the 1st and 2nd panels are consistent with the noise behaviour and plausibly not associated with the \hi\ emission or absorption of JW100. These peaks are not included in the \hi\ mask that is used to make the \hi\ map and the global profile.

The \hi\ and \co\ emissions in the channel maps (4-8th panel) are not exactly co-located, or anti-correlated, which suggests several possibilities. It could be because of a high rate of ionisation of the \hi\ gas by the young massive/OB stars, or due to an efficient conversion of \hi\ to \hh.

\ha\ emission is extended over larger velocity range than \hi\ and \co. The systemic velocity of JW100 is $\sim$ 18500 \kms (z= 0.06189, \citealt{Poggianti2019}). And based on the TF relation, we would expect a rotational velocity of 400 km/s. Therefore, due to pure rotation of the galaxy, we would expect gas emission at velocities in the range of 18100-18900 km/s. Indeed, the channel maps show the maximum recession velocity of 18800 km/s, consistent with the rotation of the galaxy. However, the lowest velocities where the \ha emission becomes visible, is roughly 17400 km/s. This difference of 1100 km/s from the systemic velocity of the galaxy can not be accommodated by the rotation of the galaxy. Furthermore, this low recession velocity is closer to the systemic velocity of the cluster, suggesting that the \ha emission is lagging behind the galaxy along its orbit through the cluster. This is observed in other GASP galaxies falling into clusters, such as JO201 in A85 \citep{bellhouse2017, bellhouse2019}.


\section{\hi depletion channels}
\label{sec:hi_dep_chnls}

 \begin{figure}
 {%
    \includegraphics[width=.5\textwidth]{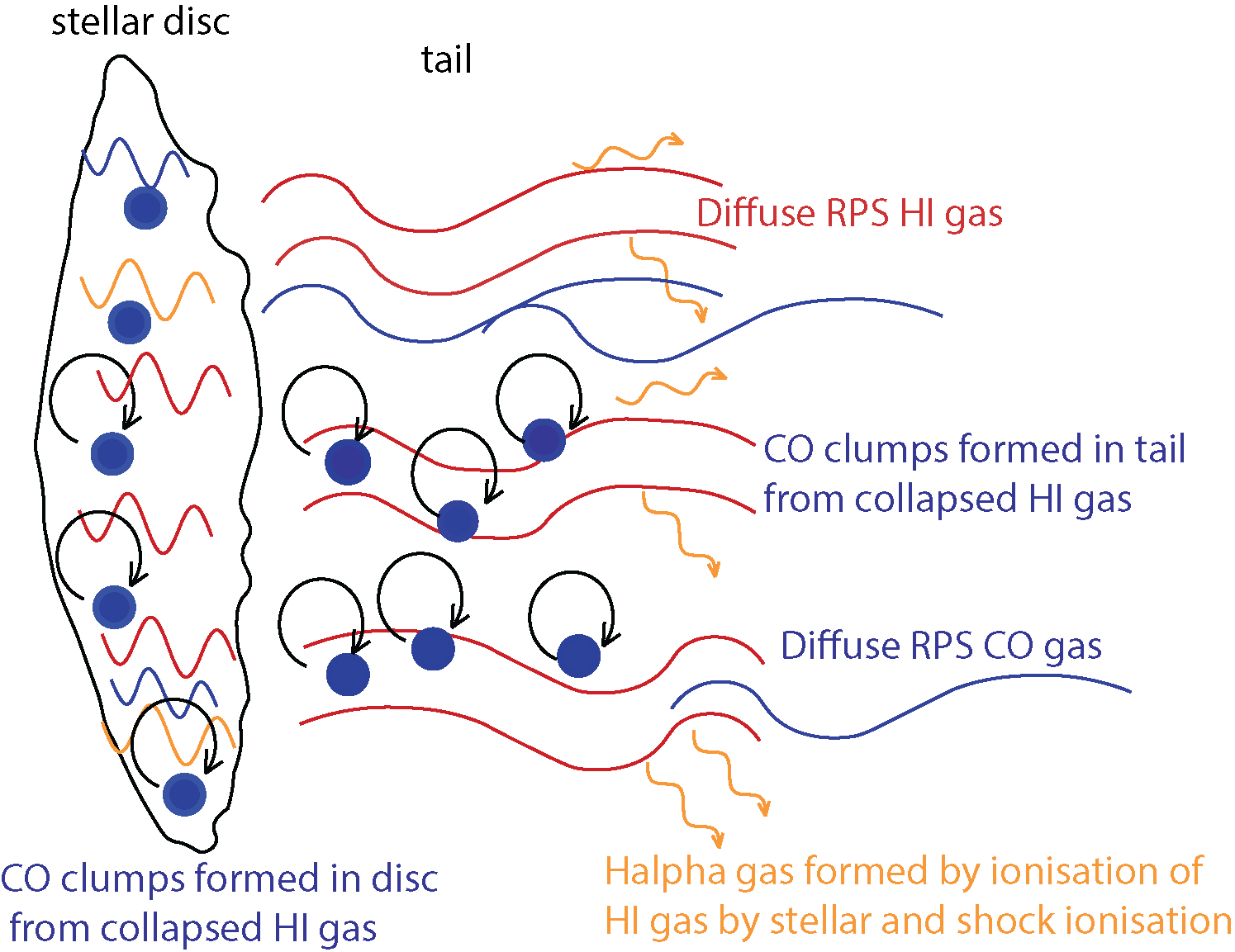}  
  } 
  \caption{Cartoon of \hi\ depletion channels. There are three principal depletion channels: ram-pressure stripping of the diffuse \hi gas, \hi to \hh conversion in the disc and tail, and ionisation of \hi gas.} 
  \label{fig:JW100_depchnls}
\end{figure}

\begin{figure}
 {%
    \includegraphics[width=.45\textwidth]{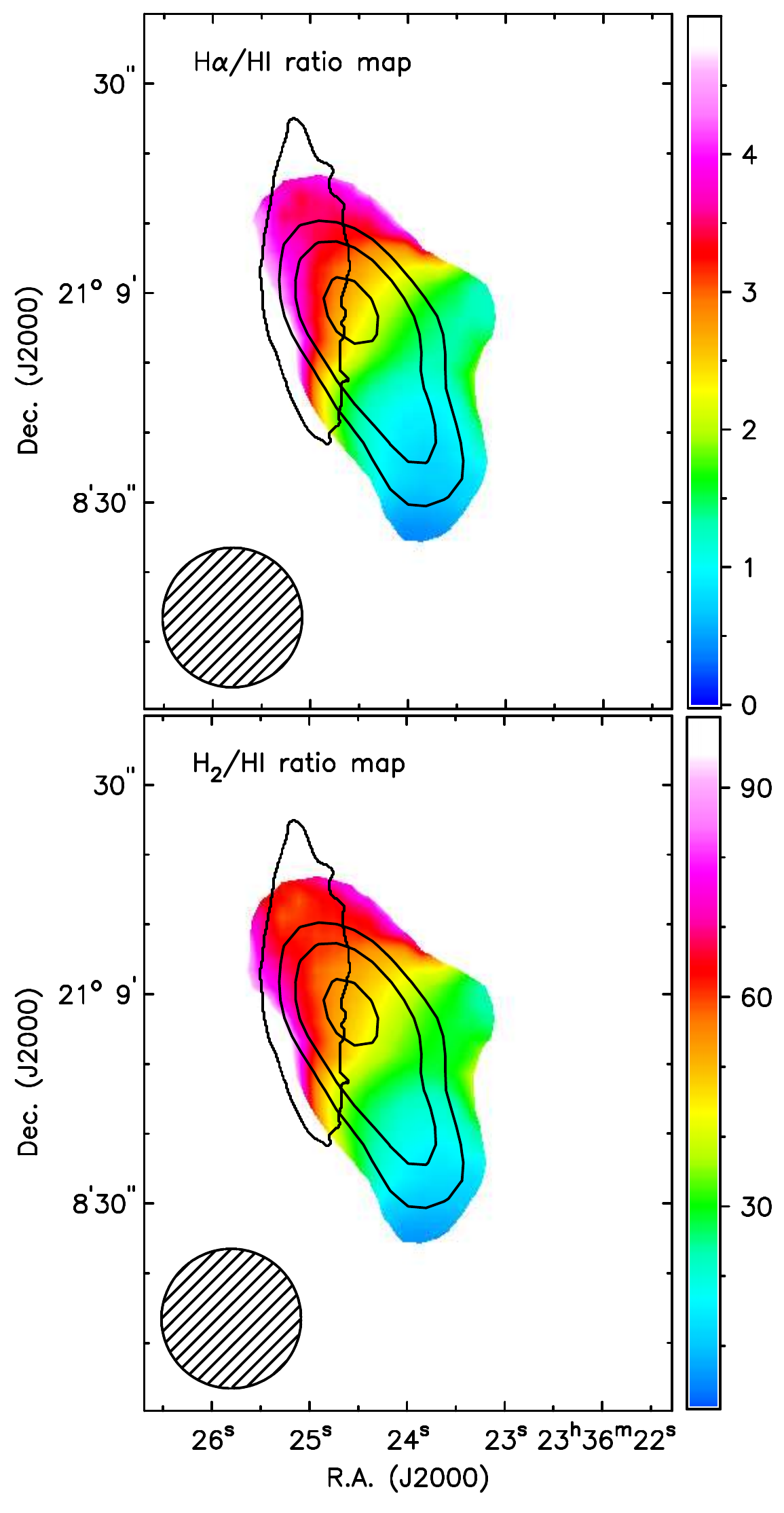} 
  } 
  \caption{\ha/\hi\ and \hh/\hi\ ratio maps of JW100. The maps are constructed by smoothing the \hi, \hh, and \ha column density maps to 20$\arcsec$ resolution and regriding the maps to the same pixel sizes (0.2$\arcsec$). The black curved contour inside represents the stellar disc. The black solid contours represent the \hi emission with contour levels (3, 4, 5,..) $\times \sigma$ or RMS. We limit the ratio map to within the 1 sigma contour of the \hi column density map. 
The beam size is shown in the bottom left. The negative gradient in both the maps show that there is an abundance of \hi over \hh or \ha along the extended tail. Considering the \mk beam size, we note that there are only 2-3 resolution elements in these ratio maps.} 
\label{fig: HA_HI_H2_dep} 
 \end{figure}
 
In what follows, we sketch a qualitative picture of possible \hi depletion channels. We estimate the relative importance of several potential \hi\ depletion mechanisms that act on a galaxy undergoing RPS. We identify three main depletion channels for \hi gas in JW100 (see schematic diagram Fig. \ref{fig:JW100_depchnls}). The relevance of each depletion channel depends on the amount of atomic or molecular gas that JW100 harboured before the onset of RPS (see Table \ref{tab:gas_masses}).

We note here already that simulations suggest that up to 50\% of the gas can be re-accreted \citep{schulz2001, vollmer2001, roediger2006, kapferer2009, quilis2017, tonnesen2019} and gas fallback has been observed in at least one galaxy \citep{cramer2021}. Our data do not evidence the presence of redshifted \hi gas. This is not surprising, as JW100 is moving at an exceptional high velocity with respect to the cluster, which makes the gas difficult to fall back onto the galaxy. Therefore, we ignore in our calculations the gas that might be falling back to the disc.

\subsection{Stripping (ram pressure)}

One of the pivotal depletion mechanisms is ram-pressure stripping of the diffuse \hi\ gas from the disc while JW100 is falling into the centre of A2626 and interacting with the hot, X-ray emitting ICM.

If we assume that the only effect of the RPS is the displacement of the initial \hi gas content, then we can infer the amount of ram-pressure stripped gas based on the mass of \hi\ gas in the tail of JW100. The mass of \hi\ gas in the tail is derived by subtracting the amount of \hi\ gas inside the stellar disc from the total \hi\ gas seen in emission. The stellar disc is derived from the MUSE observations \citep{Poggianti2019}. Hence, the total \hi\ mass in the tail is: (2.8 $\times$ 10$^{9}$ -- 0.8 $\times$ 10$^{9}$) \msun = 2 $\times$ 10$^{9}$ \msun. Thus, at least 70\% of the \hi\ gas has been removed from the disc.

\begin{table}
\renewcommand{\arraystretch}{1}
\setlength{\tabcolsep}{8pt} 
\centering
\caption{\textbf{Expected and observed gas masses in JW100 } }
\centerline{
\footnotesize{
\begin{tabular}{|c|cc|ccc|}\hline
\multicolumn{1}{|c}{} & \multicolumn{2}{|c}{Expected mass$^{a}$} & \multicolumn{3}{|c|}{Observed mass} \\
 \hline
\multicolumn{1}{|c}{Gas phase}      &
\multicolumn{1}{|c|}{value}  & 
\multicolumn{1}{c}{range} &
\multicolumn{1}{|c|}{total} & 
\multicolumn{1}{c|}{disc} &
\multicolumn{1}{c|}{tail} \\
\hline
 \multicolumn{1}{|c}{} & \multicolumn{5}{|c|}{(in units of 10$^{9}$ \msun)}  \\

\hline
\hi &  19 & 8.6-42$^{a}$ &  2.8 & 0.8 & 2.0 \\ 
\hh &  3.5 & 1-10$^{b}$ &  6-25$^{c}$ & 5-24$^{c}$ & 1.9 \\ 
Stellar & & & 320 & & \\
\hline
\end{tabular}
}
}
\begin{flushleft}
    \footnotesize $^{a}$ based on scaling relations from \cite{denes2014},$^{b}$ based on \cite{saintonge2017},$^{c}$ based on \cite{Moretti2020a}, the range corresponds to different values of $\alpha_{CO}$ varying from 0.9 (\citealt{amorin2016}) to 4.3 (\citealt{bolatto2013}).
    \end{flushleft}
\label{tab:gas_masses}
\end{table}

\subsection{\hi to \hh conversion}

For field galaxies in the xGASS sample \citep{catinella2018}, depending on their stellar masses (9.18 $\leq$ log\mstar/\msun $\leq$ 11.2), the \hi-to-\hh mass ratio varies in the range of 0.03-1.3 . In the Fornax cluster, however, \cite{Loni2021} found M$_{\hh}$/\mhi in the range of 0.1-10 for galaxies with stellar masses 9 $\leq$ log\mstar/\msun $\leq$ 11 \citep{Zabel2019, Loni2021}. In the case of JW100, M$_{\hh}$/\mhi is 8.9, which is higher than the field galaxies in the xGASS sample, but near the high end of the M$_{\hh}$/\mhi range for Fornax galaxies \citep{Zabel2019, Moretti2020b, Loni2021}. From this observation, \cite{Moretti2020a} inferred an efficient conversion of atomic to molecular gas in JW100, reinforcing the notion that \hi to \hh conversion is a significant \hi depletion channel.

We aim to investigate the relative efficiency of \hi-to-\hh conversion in the disc and the tail of the JFG JW100. In Table \ref{tab:gas_masses}, we summarise the estimated amount of \hi and \hh in the disc and the tail with the uncertainties. In the discussion that follows, we assume the maximum masses from Table \ref{tab:gas_masses}. This may introduce some uncertainty in the final calculation, but the overall conclusion of our considerations remains unaffected.

\subsubsection{\textbf{\hi to \hh conversion in the disc}}

The current amount of \hi in the disc is an order of magnitude less than the expected amount of \hi with which the galaxy entered into the cluster (see Table \ref{tab:gas_masses}). The missing \hi in the disc is either stripped from the disc or locally converted into \hh. However, the amount of \hh in the disc is roughly a factor 5 larger than the estimated amount of \hh before the onset of RPS. The amount of \hh gas that the disc gained is: observed \hh mass in the disc minus expected \hh mass which results in 14 $\times$ 10$^9$ \msun. Therefore, the excess of \hh in the disc might have formed from \hi that was present in the disc before the onset of RPS. Thus, if all the \hh in the disc is formed from \hi gas then (14/42=) 1/3 of \hi is converted into \hh in the disc.

\subsubsection{\textbf{\hi to \hh conversion in the tail}}

A significant fraction of the \hi gas in the disc is pushed out into the tail where it can also be converted into \hh. Interestingly, combining LOFAR, MeerKAT, and JVLA radio continuum observations of JW100, \cite{ignesti2022} estimated the strength of the magnetic field which could be strong enough to shield the stripped ISM gas from the ICM \citep{Mueller2021}. The draping of the magnetized plasma around the tail is possibly preventing thermal conduction and hydrodynamic turbulence by the hot ICM, enabling the stripped \hi\ gas to cool and condense into molecular clumps (e.g., \citealt{Sparre_2020, Ge_2021, Mueller2021}). The fact that we observe ongoing star formation associated with these clumps suggests that these clumps fragment further into giant molecular clouds \citep{Poggianti2019, Moretti2020a}.

Following the calculation in the previous subsection, the \hi gas that is pushed out of the disc into the tail, eventually into the ICM is : expected amount of \hi mass minus the amount of \hi in the disc minus the amount of \hi converted into \hh in the disc which results in 27 $\times$ 10$^9$ \msun. Thus, by assuming that \hh in the tail is formed from \hi,  we infer that 1.9/27 $\sim$ 7\% of the diffuse \hi gas is converted into \hh gas in the tail. In the previous subsection, we concluded that 1/3 of the \hi is locally converted into \hh. Hence, the conversion of \hi into \hh in the disc is several times more efficient than in the tail. Thus, 30+7=37\% of the \hi is converted into \hh and 60\% is blown out of the galaxy.

\subsubsection{\textbf{\hh vs. \hi ratio map}}
We have calculated the spatially resolved atomic to molecular gas fraction for both the disc and the tail. The relative amounts of both \hi and \hh in the disc and the tail are based on the stellar disc defined by the MUSE observations \citep{Poggianti2019, Moretti2020b}.
The bottom panel in Fig. \ref{fig: HA_HI_H2_dep} presents the ratio map of \hh\ vs \hi\ surface mass density (both measured in units of \msun/pc$^{2}$). The map is constructed by smoothing the \hi and \hh column density maps to 20$\arcsec$ resolution and regriding the maps to the same pixel sizes (0.2$\arcsec$). Though we are limited by the resolution of the \hi\ data, evidently, there is a negative gradient along the tail, implying there is more \hi\ compared to \hh\ towards the tip of the tail.

\subsection{\hi to \ha conversion}

Ionisation of \hi\ owing to radiation from young stars or shock ionisation is another depletion channel for \hi gas. We have investigated the spatially resolved ionised to atomic gas fraction for both the disc and the tail of JW100 using the same strategy we used for the molecular to atomic gas fraction. The top panel in Fig. \ref{fig: HA_HI_H2_dep} the shows the ratio map for \ha\ vs \hi\ surface mass density (surface mass density is measured in units of \msun/pc$^{2}$ and \ha\ emissivity in erg/s/cm$^{2}$/pc$^{2}$) for JW100. In this case also, there is a gradient along the tail, implying that the ionisation efficiency of neutral gas is decreasing along the tail.  The lower limit of \ha\ gas estimated from the MUSE observations is 4.6 $\times$ 10$^{6}$ \msun. Thus, we note that the \hi mass in the tail is an order of magnitude more than the ionised gas mass from which we infer that the ionisation of neutral gas plays a minor role in the depletion of the neutral atomic gas.


\section{Summary}
\label{sec:summary}

In this work, we have presented \mk\ \hi\ observations of the JFCGs in the A2626 galaxy cluster. Among the six JFCGs as identified by \cite{poggianti2016} from their optical morphologies in WINGS B-band images,  only JW100 is identified as a JFG from its optical and \hi\ morphologies. JW103 is probably a ram-pressure stripped galaxy at a later stage of stripping. Both JW100 and JW103 are moving at a very high similar velocity (cz $\sim$ +1800 \kms, $\sim 3\sigma_{cl}$ ) with respect to the ICM of A2626, suggesting that these galaxies are entering as a group, confirmed by \cite{HealySS2021}. Both of these galaxies reside close to the X-ray emitting cluster core, thus in the most favourable conditions for RPS. In our upcoming work, taking advantage of these \mk \hi observations, we expect to identify more RPS galaxies in the A2626 volume that could not be identified from their optical morphologies.

- The four JFCGs namely JW98, JW99, JW101, JW102 have significant \hi\ masses ($\sim$ 10$^{10}$ \msun). Their \hi\ imaging reveals warps, morphological and kinematic asymmetries, possible tidal interactions and no convincing signature of RPS. 

- JW103 is barely detected in \hi\ and very \hi\ deficient (\mhi =  10$^{9}$ \msun). While in the optical image of JW103, we observe blue tentacles towards the west, we do not detect any \hi\ tails in the direction of the optical tail, possibly because the \hi\ emission is below the detection limit. Interestingly, we see a radio continuum tail in the direction of the optical tail, which is possibly due to advection of relativistic electrons by the ram pressure wind \citep{ignesti2022}.

- We investigated \hi\ vs stellar masses for JFCGs compared to a reference sample of galaxies from the PP, UMa, A2626 and ALFALFA surveys of field galaxies \citep{maddox2015}. All JFGs have lower \hi\ masses for their stellar masses compared to most field or cluster galaxies while other JFCGs are \hi rich. We then plotted SFR vs stellar masses for the JFCGs, the reference sample galaxies and the SFMS relations. None of the JFCGs and previously studied GASP JFGs stand out with respect to the SFMS of \cite{cluver2020}.

- Finally, when we compared the \hi\ content vs the SFR in the JFCGs and the reference sample, all five JFGs stand out in the relation because of the lower \hi\ masses for their SFRs. It means that these galaxies are \hi stripped but not yet quenched.

- \hi\ observations of JW100 reveal an extended \hi\ tail towards the south-west up to $\sim$ 50 kpc  from the stellar body of the galaxy. Comparing the \ha, \co, and \hi\ channel maps in JW100, we detect \ha\ emission in the northern part of the RPS tail with no \co or \hi\ counterpart (but where X-ray emission is present, \citealt{Poggianti2019}) possibly because of complex ISM-ICM interaction due to prolonged RPS. Moreover, both \ha\ and CO are anti-correlated with \hi, hinting at an efficient conversion of \hi\ to  \hh\ in the tail. We hypothesize that the draping of the magnetic plasma around the tail is preventing thermal conduction and hydrodynamic turbulence by the hot ICM, enabling the stripped \hi\ gas to cool and condense into \hh\ clumps (e.g. \citealt{Mueller2021}).

- The spatially resolved \ha/\hi\ and \hh/\hi\ ratio maps of JW100 show a gradient along the tail, implying there is more \hi\ compared to \hh\ or \ha\ toward the tip of the extended tail. We estimated the relative importance of the different depletion mechanisms of the \hi\ gas in JW100 and we identify three main depletion channels:  RPS removal, conversion of \hi\ to \hh, and ionisation of the \hi. We found that both RPS and \hi-to-\hh conversion are significant depletion channels. \hi-to-\hh conversion is more efficient in the disc than in the tail.

\section{Data Availability}

The data underlying this article will be shared on reasonable request to the corresponding author.

\section*{Acknowledgements}

We thank Tom Jarett for kindly providing the SFRs and setllar masses for the A2626, PP, and UMa galaxies from WISE observations. TD thanks Pooja Bilimogga for providing \hi data on PP and UMa galaxies. This paper makes use of the MeerKAT data (Project ID: SCI-20190418-JH-01). The MeerKAT telescope is operated by the South African Radio Astronomy Observatory, which is a facility of the National Research Foundation, an agency of the Department of Science and Innovation. MV acknowledges the Netherlands Foundation for Scientific Research support through VICI grant 016.130.338 and the Leids Kerkhoven-Bosscha Fonds (LKBF) for travel support. Based on observations collected at the European Organization for Astronomical Research in the Southern Hemisphere under ESO programme 196.B-0578, this project has received funding from the European Research  Council (ERC) under the European Union's Horizon 2020 research and innovation programme (grant agreement No. 833824). JMvdH acknowledges support from the European Research Council under the European Union's Seventh Framework Programme (FP/2007-2013)/ERC Grant Agreement nr. 291531. JH has received funding from the European Research Council (ERC) under the European Union’s Horizon 2020 research and innovation programme (grant agreement No 882793/MeerGas). AI acknowledges the Italian PRIN-Miur 2017 (PI A. Cimatti). Y.J.  acknowledges financial support from CONICYT PAI (Concurso Nacional de Insercion en la Academia 2017) No. 79170132 and FONDECYT Iniciacion 2018 No.  11180558.



\bibliographystyle{mnras}
\bibliography{reference} 




\bsp	
\label{lastpage}
\end{document}